\documentclass[11pt]{article}
\usepackage{jheppubmod}
\usepackage{color}
\usepackage[showonlyrefs]{mathtools}
\usepackage{array}
\usepackage{amssymb}
\usepackage{amsmath}
\usepackage{slashed}
\usepackage{epstopdf}
\usepackage{cite}

\author[a]{Jan de Boer}
\author[b]{Bianca Dittrich}
\author[c]{Astrid Eichhorn}
\author[d]{Steven B. Giddings}
\author[e]{Steffen Gielen}
\author[f]{Stefano Liberati}
\author[g]{Etera R. Livine}
\author[h]{Daniele Oriti}
\author[i]{Kyriakos Papadodimas}
\author[j]{Antonio D. Pereira}
\author[k]{Mairi Sakellariadou}
\author[l]{Sumati Surya}
\author[m]{Herman L. Verlinde}

\affiliation[a]{Institute for Theoretical Physics, University of Amsterdam,
1090 GL Amsterdam, The Netherlands}
\affiliation[b]{Perimeter Institute, 31 Caroline Street North, Waterloo, ON, N2L 2Y5, CAN}
\affiliation[c]{CP3-Origins, University of Southern Denmark, Campusvej 55, DK-5230 Odense M, Denmark}
\affiliation[d]{Department of Physics, University of California, Santa Barbara, CA 93106, USA}
\affiliation[e]{School of Mathematics and Statistics, University of Sheffield,
Hicks Building, Hounsfield Road, Sheffield S3 7RH, United Kingdom}
\affiliation[f]{SISSA, International School for Advanced Studies, via Bonomea 265, 34136 Trieste, Italy
IFPU, Institute for Fundamental Physics of the Universe, via Beirut 2, 34014 Trieste, Italy
INFN, Sezione di Trieste, via Valerio 2, 34127 Trieste, Italy}
\affiliation[g]{Univ de Lyon, ENS de Lyon, Laboratoire de Physique, CNRS UMR 5672, Lyon 69007, France}
\affiliation[h]{Arnold Sommerfeld Center for Theoretical Physics,
Ludwig-Maximilians-Universität München
Theresienstrasse 37, 80333 München, Germany}
\affiliation[i]{Theoretical Physics Department, CERN, CH-1211 Geneva 23, Switzerland}
\affiliation[j]{Institute for Mathematics, Astrophysics and Particle Physics (IMAPP)
Radboud University, Heyendaalseweg 135, 6525 AJ Nijmegen,The Netherlands, Instituto de Fısica, Universidade Federal Fluminese,
Av. Litoranea s/n, 24210-346, Niteroi, RJ, Brazil}
\affiliation[k]{Physics Department, King’s College London, Strand, London, WC2R 2LS, UK}
\affiliation[l]{Raman Research Institute, Sadashivanagar, Bangalore 560080, India}
\affiliation[m]{Physics Department, Princeton University, Princeton, NJ 08544, USA}

\title{
Frontiers of  Quantum Gravity: shared challenges, converging directions\footnote{Contribution to Snowmass 2021}
}

\abstract{Understanding the quantum nature of spacetime and gravity remains one of the most
ambitious goals of theoretical physics. It promises to provide key new insights into  fundamental
particle theory, astrophysics, cosmology and the foundations of physics.  Despite this common goal, the
community of quantum gravity  researchers is
sometimes seen as divided into sub-communities working on different, mutually exclusive approaches. In practice however, recent
years have shown the emergence of common techniques, results and physical ideas arising from different
sub-communities, suggesting exciting new prospects for collaboration and interaction between traditionally distinct
approaches. In this White  Paper we discuss some of the common themes which have seen a growing interest from
various directions, and argue that focusing on them will help
the quantum gravity community as a whole towards shared objectives.}

\notoc

\begin{document}

\maketitle
\newpage
\section*{}

\noindent
{\bf Introduction and overview} --- Quantum gravity (QG) promises to shake the very foundations of our understanding of
nature by redefining its pillars, namely our current notions of space, time and matter. This is potentially of immense physical relevance and could  lead to   new
phenomenology in gravitational physics, astrophysics, particle physics and cosmology, in addition to resolving the
mysteries of black hole physics  and the very early Universe.

    Several approaches to QG have developed over the last few decades, starting from diverse  and sometimes even contradictory  assumptions and key ingredients.  Nevertheless, while retaining their  distinct character, they  sometimes converge, both in broad techniques as well as results.  Some pertinent examples of common physical principles and techniques in diverse approaches are:
    (i) holography as well as  topological field theory, common to  string theory
    and spin foams, (ii)
    the nature of low energy effective field theories studied in both  string theory (swampland conjecture) and  asymptotic safety, (iii) discreteness via  dynamical graphs and lattices in spin foams, tensor models, group field theory, causal dynamical triangulations and causal set  theory, (iv) spin network states as the  fundamental degrees of freedom of  quantum spacetime in canonical loop quantum gravity, spin foams,  tensor models and group field theory, (v)  the central role played by causality in causal dynamical triangulations and causal set theory, (vi) the use  of tensor networks in AdS/CFT, group field theory and loop quantum gravity, (vii) the role of dynamical topology in string theory, tensor models, causal sets and  group field theory, (viii) the emergence of   noncommutative geometry  in string theory, loop quantum gravity, spin foam models  and perturbative quantum gravity, (ix) the use of renormalization group techniques and the search for corresponding fixed points in asymptotic safety, spin foams, group field theories, tensor models and causal dynamical triangulations.  
    
    All of these approaches moreover share common physical goals and ideas, even when they do not use the same technical tools. These include finding the gauge-invariant observables of quantum gravity,  understanding the fundamental nature of black hole physics, investigating the possible emergence of spacetime from more fundamental degrees of freedom and understanding the relationship between entanglement and geometry, finding the quantum origins of cosmological initial conditions and dark energy, and understanding  the scattering amplitudes of the theory.

This is not to say that there are no contradictions -- often the fundamental principles are vastly different as are the various perspectives they advocate. However, what is striking about recent developments in these approaches is not only a convergence to  common goals (search for observables and the S-matrix, recovering Lorentz invariance, explanations of black holes, understanding of cosmology, etc.) but also the use of similar techniques. Increasing  the  cross talk between these communities will  therefore be of great advantage to the endeavor of quantum gravity even if there is no explicit  convergence in fundamental perspectives. 

Attempting to answer the   broad questions from these    different perspectives  chips away at and exposes  the many facets of the ``problem of quantum gravity".
Each of these questions has a long tradition which  goes beyond the  specific formalism. Some of these have been successfully tackled from many perspectives, an important example being the Entropy-Area relation for quantum black holes. The insights gained along one route have been helpful in pursuing other routes, and these sometimes  coalesce and  reinforce each other. Conversely, it also is true that contrasting different perspectives and techniques used in the diverse approaches is helpful in highlighting the nature of the  stumbling blocks in a given approach,  leading  to novel ideas or tools -- either borrowed, or freshly forged. 
Thus, increased dialogue and
cross-fertilization between different quantum gravity approaches benefits the whole endeavor, and should be actively encouraged. 

Below we overview a number of key issues and approaches in quantum gravity, ranging from the more conceptual to the more physical. For each of them, we highlight recent results obtained in one or the other of the existing quantum gravity approaches. By doing so, we emphasize how this progress has come from different directions and using different tools, but also that there is an increasing convergence of methods and results. 
We are confident that this convergence will only intensify further in the coming years, leading  to paradigm shifts and new insights of wide relevance to physics as a whole.\\

\noindent{\it disclaimer:} What follows is a kaleidoscopic, informal, and incomplete overview. Most papers on quantum gravity are written from a particular perspective and rely on various explicit and implicit assumptions. We will usually not explicitly state those assumptions in this white paper, nor do we attempt to include any sort of value judgment of those assumptions.\\

\noindent{\bf Holography}\\
The idea of holography \cite{tHooft:1993dmi, Susskind:1994vu} has become increasingly central over the last 30 years, starting from black hole physics, as a key element in understanding spacetime geometry and gravitational physics at a more fundamental level. The most developed example of holography in the quantum gravity context is by far the proposed AdS/CFT correspondence \cite{Maldacena:1997re, Gubser:1998bc, Witten:1998qj}. This correspondence has led to important insights on various questions in Quantum Gravity. For instance, the understanding of an exact CFT duality to gravity in anti-de Sitter spacetime would settle, as a matter of principle, the question of unitarity during black hole evaporation and, for example, could allow a counting of black hole microstates successfully reproducing the Bekenstein-Hawking entropy\cite{Strominger:1996sh}.

Most case studies of AdS/CFT involve semiclassical gravitational bulk theories related to string theory and supersymmetric boundary CFTs. However, it has been  suggested that holography may follow directly from general quantum gravity considerations, and specifically gauge invariance/constraints (see e.g.~\cite{Marolf:2008mf,Jacobson:2012ubm,Jacobson:2019gnm,Giddings:2020usy,Chowdhury:2021nxw} for discussions). Moreover, preliminary ideas for a  bulk/boundary correspondence have been formulated for de Sitter and flat boundary conditions. Holographic behavior has  been studied more generally in other approaches to quantum gravity like spin foams, loop quantum gravity and group field theory~\cite{Bonzom:2015ans,Dittrich:2018xuk,Dittrich:2017hnl,Chen:2021vrc, Colafranceschi:2021acz}. Here the focus has been on defining the bulk/boundary maps at the level of the quantum states,
and  on the subsequent identification of the conditions under which these maps become holographic. 

It is widely believed that holographic behavior can be traced back to the entanglement structure of the fundamental degrees of freedom living on bulk and boundary. Correspondingly, quantum information techniques for controlling entanglement in quantum many-body systems have become popular in the study of holographic behavior in (quantum) gravitational systems, notably tensor networks ~\cite{Pastawski:2015qua,Miyaji:2016mxg,Bao:2018pvs,Han:2017uco}. These  tensor network techniques have  themselves become  crucial to  quantum gravity approaches ~\cite{Dittrich:2014mxa,Colafranceschi:2021acz} which  use  spin networks,  themselves a special case of (symmetric) tensor networks. 

There is therefore evidence that holography could  play a central role in quantum gravity beyond AdS/CFT or string theory, and that more general quantum gravity considerations are needed to understand its  origins.\\

\noindent{\bf Entanglement/geometry correspondence}\\
Following the work on holographic behavior in a quantum gravity context, entanglement has been suggested to be the material that threads spacetime and geometry into existence.~\cite{VanRaamsdonk:2010pw,Bianchi:2012ev,Maldacena:2013xja}
A number of measures of entanglement have been shown to admit a geometric interpretation, both in the AdS/CFT context~\cite{Swingle:2009bg,Ryu:2006ef,VanRaamsdonk:2016exw} (see also \cite{Bousso:2022ntt,Faulkner:2022mlp} for a summary of some recent developments) and beyond, and quantum gravity formalisms in which spacetime is emergent from non-spatiotemporal fundamental quantum entities are increasingly focusing on their entanglement properties to reconstruct geometry.  

This  direction has been pursued in canonical loop quantum gravity and group field theories \cite{Colafranceschi:2020ern, Chirco:2021chk}, where the entanglement/geometry correspondence
is  explicitly  realized in terms of discrete geometry. 
Tensor network techniques have also been seen to be  central to these calculations. These entanglement entropy calculations are regularized by using a UV cut-off. In  causal set theory, there is a natural covariantly defined UV cut-off, and while the expected  Entropy-Area  relationship is satisfied away from the deep UV regime, this is modified to an  Entropy-Volume relation in the extreme UV \cite{Yazdi}. \\

\noindent{\bf Renormalization group} \\ 
The Renormalization Group (RG),  with its  corresponding concepts of universality, (quantum)
scale transformations and coarse-graining, is currently emerging as a focal point for different quantum gravity approaches, most
notably asymptotically safe gravity \cite{Bonanno:2020bil}, spin foam models (seen from a lattice gauge theory perspective) \cite{Dittrich:2011zh,Dittrich:2014mxa,Bahr:2016hwc}, dynamical triangulations
\cite{Ambjorn:2020rcn}, tensor models \cite{Eichhorn:2018phj} and group field theories (also seen as an alternative implementation of renormalization for spin foam models)\cite{Carrozza:2016vsq}.

Different incarnations of RG flows are set up in quantum gravity: in settings with (auxiliary) backgrounds, local coarse graining setups similar to those in gauge theories are used \cite{Reuter:2012id}; in background-independent settings, a more abstract notion of coarse graining is implemented which uses the number of degrees of freedom as an RG scale \cite{Bahr:2009qc,Dittrich:2014ala,Eichhorn:2018phj,Carrozza:2016vsq}; in holographic RG flows, there is an interplay between bulk and boundary degrees of freedom \cite{Skenderis:2002wp}; 
finally, in causal sets, a ``renormalization" of parameters occurs in a series of cosmic bounces within a cosmic evolution \cite{Martin:2000js}. 

The RG provides a common language that helps to establish links between distinct approaches: questions of universality, the
continuum limit, and the fate of symmetries take center stage here. In particular, a universal continuum limit may emerge from these distinct approaches -- much like a universality class in statistical physics unites microscopically distinct models under certain conditions. To discover whether or not there are several universality classes in quantum gravity, a comparison of critical exponents, initiated, e.g., in \cite{Biemans:2016rvp}, is necessary and will become feasible once computations have advanced to the required precision.

Moreover, the RG flow acts as a
bridge between the microscopic QG regime and macrophysics, where observations are possible. The interplay of QG with matter fields is being investigated within such a framework, connecting QG with high-energy physics in (beyond) Standard Model
settings \cite{Shaposhnikov:2009pv,Eichhorn:2018whv,Eichhorn:2020sbo} which also links to ongoing efforts in particle physics  such as the search for dark matter \cite{Reichert:2019car,Eichhorn:2020kca}. In spirit, the program of asymptotically safe matter-gravity models resonates with the swampland program in string theory (for explicit studies of swampland conjectures in asymptotic safety, see \cite{Basile:2021krr}). In practice, studies indicate that the UV quantum scale symmetry underlying asymptotic safety could have high constraining power that could fix many properties of the matter sector at low energies \cite{Shaposhnikov:2009pv,Eichhorn:2018whv,Eichhorn:2020sbo}, making the paradigm testable without requiring direct access to the Planck scale.  

Another important point of contact with other approaches is to understand whether and how asymptotic safety is realized as a property of {\it physical} scattering amplitudes, or related to other properties, by investigating gravitational scattering in the ultraplanckian regime.\cite{Giddings:2010pp}\\

\noindent {\bf Causality and analyticity}\\
Causality is an essential ingredient of relativistic physics, and  a guiding principle in quantum
field theory; together with analyticity, it often provides powerful constraints. Classically the causal structure of a causal spacetime is a partially ordered set which encodes all but the conformal degree of freedom.\cite{hkm,malament} This suggests that causality may be one of the most rudimentary principles in
nature.  Causality is also important in constructing the covariant observables of QG if these are assumed to be space-time in character \cite{forks}. Various proposals for the resolution of the black hole information paradox and the closely related firewall paradox\cite{Mathur:2009hf,Almheiri:2012rt} rely on the existence of small non-local effects in quantum gravity. It would be interesting to understand how to reconcile small departures from locality with the expected constraints from causality.

QG approaches
which incorporate causality in a fundamental way thus give us a vantage point not easily afforded by other
approaches. These include causal dynamical triangulations\cite{cdt}  as well as causal set theory\cite{cst}, both of
which are discrete approaches. In the former, discreteness is used as a tool whereas in the latter it is assumed to be 
fundamental, but without violating local Lorentz invariance\cite{dwlb}. Local causal structures are also implemented in Lorentzian spin foam models and group field theories, in the sense of simplicial quantum gravity. 

Evaluating the Lorentzian path integral involves a number of challenges shared across approaches. One question is which kind of configuration to allow in the path integral, i.e., whether to implement some strong notion of micro-causality or not~\cite{Ambjorn:1998xu,Asante:2021phx}. Another question is how to effectively evaluate a path integral over highly oscillating amplitudes. Recent work~\cite{Feldbrugge:2017kzv,Han:2020npv,Asante:2021phx} spanning various approaches  has employed Picard--Lefschetz methods based on a deformation of the integration contour into complexified configuration space. This brings up yet another question, namely which kind of complex metrics one should consider~\cite{Louko:1995jw,Kontsevich:2021dmb,Witten:2021nzp,Lehners:2021mah}.

The Lorentzian path integral  studied in various approaches can give us concrete insights into the broad path integral framework for quantum gravity.  Recent works employing a semi-classical approximation hint at an account for unitary evolution of evaporating black holes, via islands and replica wormholes~\cite{Penington:2019kki, Almheiri:2019qdq,Marolf:2021ghr}.  Causality and analyticity also play key roles in modern studies of gravitational scattering amplitudes, and their relation to gauge theories\cite{Giddings:2011xs,Bern:2019prr} as well as in the study of gravity in AdS via boundary CFT correlators and the conformal bootstrap program\cite{Hartman:2022zik}.\\

\noindent {\bf Topology change}
\\
In interplay with the question of which kind of causality condition to impose, one can ask whether to allow topology change during time evolution.  To freeze  spatial topology is to make a choice  from the myriad of  3-dimensional spatial topologies that exist, each  characterized  by its  mapping class  group which in turn gives rise to a countable number of  inequivalent quantizations \cite{SorkinSurya}.  This suggests   that the  quantum gravitational path integral  should  include a sum over topologies. However,  Lorentzian topology change can lead to  strange causality conditions: one either has to let go of causality itself, or alternatively allow for isolated Morse type degeneracies \cite{Geroch,Borde}.  These questions are considered, at a fundamental level,  in a broad range of continuum and discrete approaches to quantum gravity. Moreover, topology changing configurations, notably wormholes, may also appear as an effective encoding of semiclassical corrections to the gravitational path integral, as recently applied to the issue of black hole evaporation \cite{Penington:2019kki, Almheiri:2019qdq,Marolf:2021ghr}.

In causal dynamical triangulations~\cite{cdt}  strong causality conditions are imposed  which exclude topology change. This mechanism has been key to ensuring a physically reasonable large-scale limit~\cite{Ambjorn:2012jv}, and in fact the suggestion is that suppressing or even excluding topology changing configurations is a necessary condition for achieving a working definition of the continuum gravitational path integral. 
In group field theories and tensor models~\cite{Oriti:2011jm, Gurau:2011xp}, on the other hand, one naturally sums over all topologies, with the  contributions from different topologies organized in a suitable field-theoretic perturbative expansion. 

These formalisms, in fact, can be seen as the modern and discrete counterpart of the third quantization idea for continuum quantum gravity \cite{Giddings:1988wv}, meant to extend both canonical geometrodynamics and the gravitational path integral approach with a sum over topologies. This third quantized formalism can also be seen at the root of the wormhole calculus  applied recently \cite{Giddings:2020yes} in the context of the calculation of black hole evaporation via the replica approach to the gravitational path integral\cite{Penington:2019kki, Almheiri:2019qdq,Marolf:2021ghr}. 

The main task, in old as well as modern incarnations of sums over topologies in quantum gravity,  becomes to control such expansions and to show that simple topologies dominate the sum, in some appropriate regime. In group field theories and random tensor models, this has been achieved within a generalized large-$N$ approximation (then further extended to multiple-scaling limits), in which specific discrete spherical topologies dominate the quantum dynamics. In other discrete approaches like causal set  theory, the path  sum includes  discrete causal geometries whose continuum approximation is a causal Lorentzian topology change. The fundamental  discreteness prevents the  Morse degeneracies of  the continuum being  realized in the causal set.  Other works investigate the stability of configurations with  topology change: the $d=2$ scalar field propagation on the trousers topology for example is known to be singular/unstable \cite{AndersonDeWitt}. Exploring these instabilities more broadly will give important insights into the role of topology change in quantum gravity. 

Apart from issues with causality conditions and stability, incorporating topology change implies interesting conceptual challenges to  the standard probability interpretation of the theory, providing another example of the overlap between research in quantum gravity and in quantum foundations. Works addressing these questions span a wide range of approaches~\cite{Hosoya:1988aa,McGuigan:1988vi,Oriti:2013aqa,Casali:2021ewu}.
\\

\noindent {\bf Symmetries and boundary charges}\\
A fundamental aspect of QG research is symmetry. This includes  possible modifications of the relativity principle within our quantum Universe, which can arise due to the graininess of space-time at the Planck scale\cite{DefLor}. The investigation of bulk symmetries and boundary charges, at both classical and quantum levels, and their holographic interplay, has led to convergence between a number of approaches\cite{Bcharge1,Bcharge2,Bcharge3,Bcharge4} and quasi-local implementations of the holographic principle\cite{Dittrich:2018xuk}.
These developments have brought to light an exciting interface between QG and extended topological quantum field
theories\cite{Dittrich:2016typ} as well as with related condensed matter models, and have provided new perspectives on coarse-graining and renormalization in QG\cite{Livine:2017xww,Cunningham}.\\

\noindent{\bf Quantum-first approach and questions of quantum foundations} \\ 
Another approach to QG is the ``quantum-first'' approach, advocated in
\cite{QFG,QGQFA}.  Here, the postulates of quantum mechanics are assumed, and the  appropriate mathematical structure on
a Hilbert space is sought guided in part by  the weak gravity ``correspondence'' limit and the properties of black holes. This is
partly motivated by the mathematical structure of quantum field theory (QFT), in terms of a net of subalgebras within  the
algebra of quantum observables on the Hilbert space.  The subsystem structure for gravity is
apparently different from that for QFT\cite{Donnelly:2016rvo,Donnelly:2018nbv,Donnelly:2017jcd,Giddings:2019hjc,Laddha:2020kvp,Chowdhury:2021nxw,Giddings:2021khn}, but can be approximately described
perturbatively, while making  contact with ideas of holography\cite{Marolf:2008mf,Jacobson:2012ubm,Giddings:2020usy}. Other important constraints come from
imposing unitary evolution in the high-energy sector when black holes are produced, and parameterizing that evolution in an effective approach\cite{NVU,BHQU}. The latter has the potential to make contact with strong-gravity observations\cite{SGnature}.

By starting with a Hilbert space and directly seeking appropriate mathematical structure, this approach differs from other approaches in which classical systems are quantized, as with quantization of classical geometry, including for example using the histories Hilbert  space \cite{historieshilbert}, and including 
other quantum gravity approaches which do not start from the quantization of continuum gravitational theories, but aim at obtaining them in some approximation. This could also provide a different pathway to a quantum theory than proposals where 
 the notions of space, time and gravity are seen as emergent from behavior of other quantum systems, as in some current approaches.  
 
 An important related question is whether quantum gravity resides within the framework of quantum mechanics, with a sufficiently general formulation of its rules,\cite{Giddings:2007ya} or whether it requires modification of quantum principles.  Here quantum gravity meets developments in quantum foundations, which are in fact directly relevant (in one form or another) to all quantum gravity formalisms. Possible points of contact are given by work on probability theories with indefinite causal structures \cite{Hardy:2006uc, Castro-Ruiz:2017gzs, Oreshkov:2011er}, and the work on the quantum counterpart of the equivalence principle \cite{Zych:2015fka} and general covariance \cite{delaHamette:2021oex} as well as the quantum measure approach in which quantum theory is viewed as a  generalization of classical stochastic theory \cite{quantummeasure}. \\

\noindent{\bf Observables} \\ 
A longstanding question across approaches, intimately  connected  to tests of  QG,
is to define observables in QG.  Gauge invariant observables in gravity cannot be
local from the point of view of the ``spacetime manifold"\cite{Torre}, and one should give a gauge invariant meaning to events and locality. An  approach going back to DeWitt\cite{DeWitt} is to define observables relationally, i.e., as correlations between dynamical fields, and describe localization in terms of suitably chosen dynamical systems used as physical frames.  There have been many recent developments in this direction\cite{Hoehn}, revealing fundamental limitations on spacetime localization\cite{GMH,DittTa}. This research ties closely to that on quantum reference frames and quantum covariance in generic quantum mechanical theories, already mentioned, and raises the fundamental issue of how physics changes when different physical frames are used\cite{Hohn:2018toe,Gielen:2020abd}.  Also, in the histories framework it is possible to define covariant observables  or ``beables",  as Bell referred to them,  as elements  in a covariant event algebra \cite{Brightwell:2002vw}. This is the approach followed in the quantum sequential growth models in  causal set theory\cite{Dowker:2010qh,Surya:2020cfm,Rideout:1999ub}. 

Another approach is to construct gauge-invariant observables by gravitationally dressing field theory observables\cite{DoGi}.  These observables begin to reveal basic gravitational noncommutativity\cite{DittTa,DoGi}, and potentially important aspects of the mathematical structure of QG\cite{Giddings:2019hjc,Giddings:2021khn}.  
A better understanding of these issues is important in understanding  the mathematical structure of QG.\\

\noindent {\bf Phenomenology} \\ 
A formidable challenge faced by QG  is the longstanding lack of strong
experimental/observational guidance. QG phenomenology\cite{AmelinoCamelia:2008qg} is a 
field of research that aims at filling this gap by extracting  theoretical predictions for new physics
in accessible energy regimes, from within different QG  approaches, and testing them via observations and
experiments in windows of opportunity where even tiny, Planck suppressed, effects could be probed. This search for tests
of QG predictions has over time developed in several directions: tests of breaking/quantum deformation of local spacetime symmetries such as local Lorentz invariance (e.g.~via high energy astrophysics observations\cite{Liberati:2013xla,Kowalski-Glikman2005}); tests of departures from locality\cite{Philpott:2008vd}  (e.g.~via tabletop experiments\cite{Belenchia:2015ake}); tests of QG induced modifications of gravitational dynamics (e.g.~observing black holes via gravitational waves\cite{Abedi:2016hgu} and/or\cite{SGnature,BHQU} very long baseline interferometry (VLBI)\cite{EHT}, or studying the consequences of dimensional flow for the luminosity distance scaling of gravitational waves\cite{Calcagni:2019ngc,Calcagni:2019kzo}); and searches for extra dimensions (e.g.~in microgravity experiments and at LHC\cite{articleAntoniadis,KRETZSCHMAR2016541}). All these avenues have required cross-field collaborations and an interplay between theoretical and experimental/observational teams. \\

\noindent {\bf Foundations of cosmology}\\ 
A deeper understanding of QG will help in bridging the gap between the Standard
Model of particle physics and the  $\Lambda$ Cold Dark Matter ($\Lambda$CDM) model of cosmology,
and its observables. In $\Lambda$CDM, dark energy, dark matter and inflation need to be added to general relativity in
order to describe the observed Universe\cite{Ade:2015xua,Ade:2015lrj}, but an exciting possibility is that all these three ingredients will ultimately find a deeper explanation within QG. QG suggests that general relativity
receives corrections which can become relevant in cosmology\cite{Berti:2015itd}. The origin of dark energy is 
tied to the quantum structure of spacetime  \cite{SorkinLambda}, and dynamical dark energy scenarios can be confronted with
observation\cite{Zwane:2017xbg}. 

In the early Universe, QG should resolve the Big Bang singularity\cite{Husain:2003ry,Ashtekar:2008ay,deCesare:2016axk,deCesare:2016rsf,Pithis:2016wzf,Gielen:2020abd} and give insights into cosmological initial conditions beyond those of $\Lambda$CDM. One possible scenario is that our expanding Universe originated in a prior contracting phase\cite{Ashtekar:2011ni,Oriti:2016qtz} or a phase characterized by quantum oscillations of the geometry~\cite{Alesci:2016xqa}. QG can also constrain early-universe dynamics; for instance, the swampland conjectures in string theory constrain inflationary models and are in tension with a cosmological constant as dark energy\cite{Heisenberg:2018yae}. 

Alternatively, QG may predict an early era of accelerated expansion in the absence of a scalar field\cite{deCesare:2016axk,deCesare:2016rsf,Pithis:2016wzf}. Given that cosmology remains strongly driven by current and near-future observations and that there are unresolved tensions in the successful $\Lambda$CDM model such as, e.g., different values of the Hubble parameter resulting from different types of observations, there is particular incentive for practitioners from all approaches to QG to work together in applying their theoretical frameworks to cosmology. For instance, QG may suggest a more refined view of dark energy, possibly suggesting its dynamical nature. 

Again, quantum gravity approaches can provide a fundamental framework in which effective field theory models of a dynamical dark energy can be embedded \cite{}, but also provide alternative mechanisms producing an effective dark energy-driven acceleration as a manifestation of underlying quantum gravity interactions \cite{Oriti:2021rvm}.\\

\noindent {\bf Linking distinct approaches to quantum gravity} \\ 
Concrete links between distinct quantum gravity approaches have the potential to speed up progress toward addressing shared physical questions as well as to solve key outstanding issues within a given approach. 

In this spirit, concrete and specific links between approaches have been investigated starting from the asymptotic-safety paradigm. 
For instance, it has been suggested that asymptotic safety could arise within the effective-field theory regime of string theory \cite{deAlwis:2019aud}, see also \cite{Basile:2021euh}. On the one hand, this would enable to connect a negative value of the cosmological constant in the UV to observationally viable values in the IR by the RG flow through an asymptotically safe regime -- thus potentially solving the challenge how to reconcile string theory with de Sitter spacetime. On the other hand, a UV completion through string theory, if it also addressed the nonperturbative strong gravity regime, would automatically unitarize asymptotic safety, which might be unitary on its own \cite{Fehre:2021eob},
but with the final verdict on this subtle question is currently outstanding.

As a second example, the search for asymptotic safety is conducted both in the continuum approach as well as in discrete approaches, interpreted as providing a regularization, first, and a more rigorous definition, then, of the continuum quantum gravitational theory (rather than a discrete alternative to it). This includes causal dynamical triangulations \cite{Ambjorn:2020rcn} as well as tensor models \cite{Eichhorn:2019hsa}. Because of the distinct nature of the approaches, systematic uncertainties can be controlled much more effectively in such a combination of approaches than in a single approach on its own.

When looking at quantum gravity approaches formulated in terms of discrete entities, on the other hand, extensive links at the level of their emergent continuum dynamics are currently harder to find, since the emergence of effective continuum physics (as expressed, for example, in the usual language of effective field theory) from the fundamental quantum dynamics is only partially under control at the moment (as opposed to kinematical reconstructions, which are available in several formalisms). 

At the more formal level, however, some links are structural and immediate; they follow from sharing the same fundamental mathematical tools. For example, spin networks states encode the fundamental quantum gravity degrees of freedom in canonical loop quantum gravity, spin foam models and group field theories, differing in how they are organized in the full (kinematical) Hilbert space of the theory. Further, spin foam models and group field theories have a quantum dynamics built on simplicial lattices and amplitudes associated to them, in close contact with simplicial gravity path integrals of lattice quantum gravity approaches. Moreover, spin foam amplitudes arise generically as perturbative Feynman amplitudes of group field theory models, while at the same time providing a covariant definition of the quantum dynamics of canonical loop quantum gravity. Another commonality is the use of Markov Chain Monte Carlo techniques for studying the analytically continued path integral via the  expectation values of covariant  observables, as the case in causal set theory and causal dynamical triangulations.  Random tensor models correspond to a purely combinatorial formulation of the lattice path integral, as in dynamical triangulations, and  the former  can also be seen as defining a generating function of the latter (at least in the Euclidean domain). 

Finally, beyond the technical challenges faced by quantum gravity approaches, either specific to a particular approach or more universal, there are several  common conceptual challenges as well that return to the forefront regularly. These include the ``problem of time", the role of non-locality in quantum gravity, black hole evolution, the question of information loss, and quantum initial conditions or spacetime singularities, among many others.
Despite adopting  technically different approaches to quantum gravity, these central questions remain at the heart  of our quest for a full theory  of quantum gravity. This in turn perhaps provides the most significant linkage between these approaches, which we feel should be nurtured. \\

\noindent {\bf Conclusions}
In this White Paper we have argued that a dedicated search for connections between distinct approaches to quantum gravity is an important way forward in the collective endeavor to find a viable theory of quantum gravity.  

We have given examples of  overlapping techniques used in different approaches -- these are already manifold, and worth strengthening. We have also discussed how various conceptual questions and perspectives overlap in  some cases, while fruitfully clashing in other cases. We have argued that it  is important  that this commonality  be built on in order to  make progress in quantum gravity.  
Beyond technical tools and formal structures, it is on conceptual and physical grounds, especially those of observational relevance,  that quantum gravity approaches need to be assessed, and it is in this spirit that they should focus on shared questions,  despite  differences in the  starting assumptions or in specific implementations. \\

\noindent {\bf Acknowledgments} A.D.P. acknowledges CNPq under the grant PQ-2 (309781/2019-1) FAPERJ under the “Jovem Cientista do Nosso Estado” program (E26/202.800/2019), and NWO under the VENI Grant (VI.Veni.192.109) for financial support. The work of SBG is supported in part by the U.S. Department of Energy, Office of Science, under Award Number {DE-SC}0011702, and by Heising-Simons Foundation grant \#2021-2819. D. Oriti acknowledges funding from DFG research grants OR432/3-1 and OR432/4-1.

\clearpage

\end{document}